\documentclass[final,5p,times,twocolumn]{elsarticle}
\usepackage{epsf, epsfig, latexsym}
\usepackage{graphicx}
\usepackage{amssymb}
\usepackage{amsmath}
\usepackage{bm}
\usepackage{bbm}
\usepackage{dcolumn}
\usepackage{multirow}
\usepackage{pifont}
\usepackage{calc}
\usepackage{color}
\journal{Physics Letters B}
\begin{document}
\begin{frontmatter}
\title{Nonlocal Condensates in Hadrons and Multiquark States}
%
%
\author[jb1]{Jong Bum Choi}
\ead{jbchoi@jbnu.ac.kr}
\author[jb1]{Eun-Joo Kim\corref{cor1}}
\ead{ejkim@jbnu.ac.kr}
\author[jb2]{Moon Q. Whang}
\author[jb2]{Jin D. Kim}
\author[jb2]{Su K. Lee}
%
%
\address[jb1]{Division of Science Education and Institute of Science Education,
Chonbuk National University, Jeonju 561-756, Korea}
\address[jb2]{Division of Science Education, Chonbuk National University, Jeonju 561-756, Korea }
\cortext[cor1]{Corresponding Author}
%
%
%
%
\begin{abstract}
%
The definition of QCD vacuum is an important issue in the description of hadronic
properties. Recent researches on the vacuum condensates have resulted in the suggestion
of in-hadron condensates which can be taken as a paradigm shift concerning the viewpoints
on the QCD vacuum. In this Letter, we will try to define the in-hadron regions and
to classify the hadronic and multiquark states. Topological classifications are
naturally introduced and by considering nonlocal measure
we can estimate the variations of dimension $2$ condensate $\langle A_{\mu}^2 \rangle$.
The calculational techniques can be easily applied to multiquark states and
are expected to be applied to the more complex nuclear states.
\end{abstract}
\end{frontmatter}
%
%
%
%
\section{Introduction}
%

Condensates are defined as the non-zero vacuum expectation values
of normal ordered products of field operators.
Since the effects of normal orderings are to remove the infinite zero-point energy
and to define the zero of energy as the energy of the vacuum state,
it is indispensable to relate the values of various condensates to the precise definition
of the vacuum state itself. Recent progresses in the research of these condensates
have resulted in the suggestion of in-hadron condensates~\cite{R1} which can be taken
as a paradigm shift concerning the viewpoints on the QCD vacuum~\cite{R2}.

For the in-hadron condensates, we assume that the strong interaction condensates
are the properties of well-defined wavefunctions of hadrons rather
than the hadron-less ground state of QCD. This viewpoint imposes the problem
of determining the shape of hadronic wavefunctions which corresponds partly
to figuring out the quark and gluon distributions~\cite{R3}.
In general, the values of hadronic wavefunctions are not constants
so that we can induce that the condensate values can vary from point to point.
Conversely, if we assume that condensates have constant values
inside hadrons, then there should be sharp boundaries
between the outside and the inside of hadrons.
Although there exist some hard-wall models, it is not natural to draw
the sharp boundary surfaces to define the in-hadron regions.
Therefore we need to consider smoothly changing functions
for the values of strong interaction condensates.

%
\section{The model : connection amplitude}
%

The motivation to introduce the in-hadron condensates is to account for the fact
that the vacuum states in a confining theory is not defined relative to the fields
in the Lagrangian but to the actual physical, color-singlet, states~\cite{R4}.
So in order to define the in-hadron regions, it is better to start with
the classifications of the color-singlet states.
Since the condensates effectively originate from $q\bar{q}$ and gluon contributions
to the higher Fock states, we have to include these states from the beginning.
Then it is convenient to represent the color-singlet states
by their number of quarks and antiquarks. For the color-singlet states
with $a$ quarks and $b$ antiquarks forming some in-hadron regions,
the difference between $a$ and $b$ is a multiple of $3$ in SU$(3)_c$ case
and is related to the baryon number of the given color-singlet state.
Because the baryon number is conserved in strong interactions, we need to categorize
the color-singlet states into the sets of correlated ones with fixed baryon number.
The correlations between different color-singlet states can be established
by tracing the changes of in-hadron regions which are represented
by the numbers of quarks and antiquarks.
With the constraint of baryon number conservation,
the methods to change the in-hadron regions are provided only
through the quark pair creation and the quark pair annihilation.
These changes of in-hadron regions can be systematically described
by introducing the notion of open sets for in-hadron regions
and the operations of union and intersection of the open sets.
The union operation corresponds to the quark pair annihilation
and the intersection operation to the quark pair creation.
Now we can write down our starting assumptions as~\cite{R5}
\newcommand{\squishlist}{
 \begin{list}{$\bullet$}
  { \setlength{\itemsep}{0pt}
     \setlength{\parsep}{0pt}
     \setlength{\topsep}{5.0pt}
     \setlength{\partopsep}{0pt}
     \setlength{\leftmargin}{1.5em}
     \setlength{\rightmargin}{1.5em}
     \setlength{\labelwidth}{1em}
     \setlength{\labelsep}{0.5em} } }
\newcommand{\squishend}{
  \end{list}  }
\squishlist
%
\item Open sets are the stable in-hadron regions.
\item The union of stable in-hadron regions becomes a stable in-hadron region.
\item The intersection between a connected stable in-hadron region
      and disconnected stable in-hadron regions is the reverse operation of the union.
\squishend
These assumptions are just those needed for constructing the topological spaces
of in-hadron regions. Let's use the notation $R_{a,\bar{b}}$
as representing the in-hadron regions formed between $a$ quarks and $b$ antiquarks.
Thus $R_{1,\bar{1}}$ represents the in-hadron regions of mesons,
and $R_{3,\bar{0}}$ and $R_{0,\bar{3}}$ correspond to those of baryons and antibaryons.
Then the simplest non-trivial set of in-hadron regions satisfying
the above three assumptions is given by
\begin{equation}
   T_{0,\bar{0}}~ = ~\{ \phi, ~R_{1,\bar{1}}, ~R_{1,\bar{1}}^{2}, \cdots, R_{1,\bar{1}}^{n}, \cdots \},
\label{eq1}
\end{equation}
where $R_{1,\bar{1}}^{n}$ denotes the in-hadron regions of $n$ mesons
which can be obtained by repeated quark pair creations.
The next baryon number $1$ topological space
for baryon-meson system is
\begin{equation}
  T_{1,\bar{0}}~ = ~\{ \phi, ~R_{3,\bar{0}}, ~~R_{3,\bar{0}}R_{1,\bar{1}},
  ~R_{3,\bar{0}}R_{1,\bar{1}}^{2}, \cdots, ~R_{3,\bar{0}}R_{1,\bar{1}}^{n}, \cdots \},
\label{eq2}
\end{equation}
and the baryon number $-1$ space for antibaryon-meson system becomes
\begin{equation}
 T_{0,\bar{1}}~ = ~\{ \phi, ~R_{0,\bar{3}}, ~~R_{0,\bar{3}}R_{1,\bar{1}},
 ~R_{0,\bar{3}}R_{1,\bar{1}}^{2}, \cdots, ~R_{0,\bar{3}}R_{1,\bar{1}}^{n}, \cdots \}.
\label{eq3}
\end{equation}
We can easily see that $R_{1,\bar{1}}$ can be multiplied at any time
without the violation of baryon number conservation,
and therefore we may omit these multiplications in our notation writing as
\begin{equation}
   T_{1,\bar{0}}~ = ~\{ \phi, ~R_{3,\bar{0}} \},
   ~~~~T_{0,\bar{1}}~ = ~\{ \phi, ~R_{0,\bar{3}} \}.
\label{eq4}
\end{equation}
In general, we have
\begin{equation}
   T_{i,\bar{j}}~ = ~\{ \phi, ~R_{3,\bar{0}}^{i}R_{0,\bar{3}}^{j},
   ~R_{3,\bar{0}}^{i-1}R_{0,\bar{3}}^{j-1}R_{2,\bar{2}}, \cdots \},
\label{eq5}
\end{equation}
where we can take $i$ as the number of incoming 3-junctions and $j$
that of outgoing 3-junctions if we represent the in-hadron region
as originating from quarks and terminating at antiquarks.
For a given space, the baryon number $B = i - j$ is conserved
and the number of boundary points encompassing the quarks and antiquarks
is reduced by $(1,1)$ pair through one union operation.
As examples we can write down~\cite{R6}
\begin{equation}
   T_{3,\bar{1}}~ = ~\{ \phi, ~R_{3,\bar{0}}^{3}R_{0,\bar{3}}, ~R_{3,\bar{0}}^{2}R_{2,\bar{2}},
   ~R_{3,\bar{0}}R_{4,\bar{1}}, ~R_{6,\bar{0}} \},
\label{eq6}
\end{equation}
and
\begin{eqnarray}
\hspace{-0.2in}
   T_{5,\bar{2}} ~=~ \{ \phi, ~R_{3,\bar{0}}^{5}R_{0,\bar{3}}^{2}, ~R_{3,\bar{0}}^{4}R_{0,\bar{3}}R_{2,\bar{2}}, ~R_{3,\bar{0}}^{4}R_{1,\bar{4}},  ~R_{3,\bar{0}}^{3}R_{0,\bar{3}}R_{4,\bar{1}},  \nonumber \\
\hspace{-0.2in}
   R_{3,\bar{0}}^{3} R_{3,\bar{3}}, ~R_{3,\bar{0}}^{2}R_{0,\bar{3}}R_{6,\bar{0}}, ~R_{3,\bar{0}}^{2}R_{5,\bar{2}},
   ~R_{3,\bar{0}}R_{7,\bar{1}}, ~R_{3,\bar{0}}^{3}R_{2,\bar{2}}^{2}, \\
\hspace{-0.2in}
   R_{3,\bar{0}}^{2}R_{2,\bar{2}}R_{4,\bar{1}}, ~R_{3,\bar{0}}R_{4,\bar{1}}^{2}, ~R_{3,\bar{0}}R_{2,\bar{2}}R_{6,\bar{0}}, ~R_{6,\bar{0}}R_{4,\bar{1}}, ~R_{9,\bar{0}} \}, \nonumber
\label{eq7}
\end{eqnarray}
and so on.

With the defined topological spaces of in-hadron regions,
we are now to try to deduce the spatial variations of in-hadron condensates.
As is well-known, there exist various kinds of condensates such as
$\langle A_{\mu}A^{\mu} \rangle$, $\langle \bar{q}q \rangle$,
$\langle G_{\mu\nu}G^{\mu\nu} \rangle$,
$\langle \bar{q}A\hspace{-0.07in}/q \rangle$ and higher dimension condensates.
All these combinations are made from gluon and quark fields,
and in order to figure out the condensate structures
it is better to consider firstly the simpler ones $\langle A_{\mu}A^{\mu} \rangle$
and $\langle \bar{q}q \rangle$. The effects of these condensates have been discussed in many ways.
One of the common conclusions is
that they are related to the generation of dynamical masses.
The dimension 2 condensate $\langle A_{\mu}^{2} \rangle$ gives rise to the gluon mass
and the momentum dependent dynamical quark mass $M(p^{2})$~\cite{R7},
and the non-zero value of $\langle \bar{q}q \rangle$ is usually held
to signal the dynamical chiral symmetry breaking
which is an efficient mass-generating mechanism resulting in the constituent quark masses.
The momentum dependences of mass function $M(p^{2})$ can be interpreted
as the position dependences in coordinate space, which, if possible,
have to be deduced from basic properties of in-hadron regions or their correlations.
Since the correlations between in-hadron regions are induced
by quark pair creations or annihilations,
it is natural for us to consider the bilocal scalar condensate of quark pair
\begin{equation}
 \langle : \bar{q}(x)U(x,0)q(0) : \rangle \equiv \langle : \bar{q}(0)q(0) : \rangle Q(x^{2}),
\label{eq8}
\end{equation}
where $U(x,0)$ represents the connection through in-hadron region.
The function $Q(x^2)$ is nonlocal and can be named
as the connection amplitude~\cite{R8}.

The form of the function $Q(x^2)$ was parameterized by introducing
vacuum distribution functions and estimated within single instanton approximation~\cite{R9}.
However, the vacuum distribution functions are not known and they have to be calculated
from QCD vacuum theory. Since we are now considering smoothly changing functions
for the values of in-hadron condensates, it is plausible to introduce
a measure $\mathfrak{M}(Q)$ to account for an intuitive picture of in-hadron regions.
First, we can assume
\begin{equation}
  \mathfrak{M}(Q)~~\mbox{decreases ~as} ~~Q~~ \mbox{increases}.
\label{eq9}
\end{equation}
This condition implies that a smaller in-hadron region is
more likely to be connected than a larger one.
The second condition can be drawn from the relation
\begin{eqnarray}
 \langle :  \bar{q}(x)U(x,y)q(y)\bar{q}(y)U(y,0)q(0) : \rangle  \nonumber \\
 = \langle :  \bar{q}(0)q(0) : \rangle^{2} Q((x-y)^{2})Q(y^{2}),
\label{eq10}
\end{eqnarray}
where one quark pair creation at $y$ divides
the original in-hadron region into two regions.
Thus we have
\begin{equation}
  \mathfrak{M}(Q_{1}) + \mathfrak{M}(Q_{2}) = \mathfrak{M}(Q_{1}Q_{2}),
\label{eq11}
\end{equation}
where $Q_{1}$ and $Q_{2}$ are taken to be independent.
This condition states that two independent in-hadron regions
can be joined to form a single in-hadron region.
Then with these two conditions, we get the solution
\begin{equation}
  \mathfrak{M}(Q)  = -k\ln\frac{Q}{Q_{0}}
\label{eq12}
\end{equation}
where $Q_{0}$ is a normalization constant and $k$ is an appropriate parameter.

The measure $\mathfrak{M}(Q)$ should be a metric function defined on the in-hadron region.
For the in-hadron region with a quark pair at position $\vec{x}$ and $\vec{y}$,
the distance function between $\vec{x}$ and $\vec{y}$ can be written as
\begin{equation}
 d(\vec{x}, \vec{y}) = |\vec{x}-\vec{y}|^{\nu}
\label{eq13}
\end{equation}
with $\nu$ being an arbitrary number. This distance function can be made metric
for points $\vec{z}$ satisfying
\begin{equation}
  |\vec{x}-\vec{z}|^{\nu} + |\vec{z}-\vec{y}|^{\nu}  \geqq  |\vec{x}-\vec{y}|^{\nu}.
\label{eq14}
\end{equation}
%
%
\begin{figure}[htb]
\begin{center}
\includegraphics[width=0.84\linewidth]{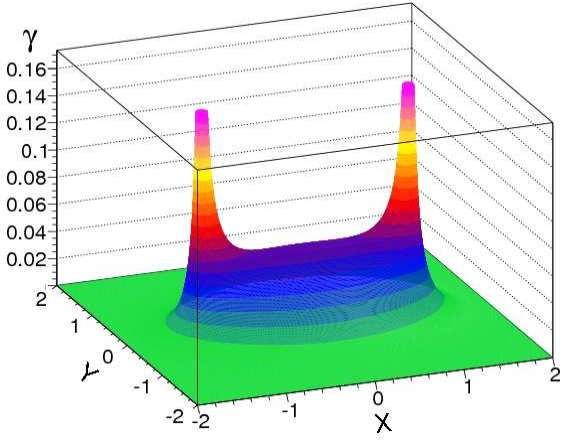}	
\caption{ Dimension 2 condensate values for a meson with
          $\beta = 1.0$ and $k = 1.0$.
          The quarks are at $(-1.175, 0)$ and $(1.175, 0)$ respectively.}
\label{fig1}
\end{center}
\end{figure}
%
%
\begin{figure}[htb]
\begin{center}
\includegraphics[width=0.84\linewidth]{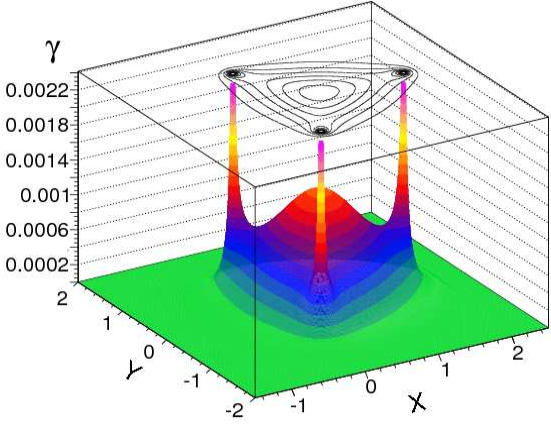}	
\caption{ Condensate values for a baryon with quarks at
          $(0, 1)$, $(0, -1)$, and $(\sqrt{3}, 0)$.
          The parameters are the same as in Fig.~\ref{fig1}
          and equi-$\gamma$ curves are shown in the upside. }
\label{fig2}
\end{center}
\end{figure}
The set of points $\vec{z}$ not satisfying this triangle inequality can be taken
as forming the inner part of the in-hadron region where it is impossible
to define a metric from the boundary points $\vec{x}$ and $\vec{y}$.
For a given value of $\nu$ we can figure out the shape of the in-hadron region,
and we may take $|\vec{x}-\vec{y}|^{\nu}$ as an appropriate measure
to deduce the form of $Q$. In order to account for various possibilities,
we need to sum over contributions from different values of $\nu$.
The lower limit of $\nu$ can be fixed to be $1$ because no point $\vec{z}$
satisfying the triangle inequality with $\nu < 1$ exists.
For a small increment $d\nu$, the product of the two probability amplitudes
for $|\vec{x}-\vec{y}|^{\nu}$ and $|\vec{x}-\vec{y}|^{\nu+d\nu}$
to satisfy the metric conditions can be interpreted
as the probability amplitude for the increased region
to be added to the inner connected region, which is out of the metric condition.
Then, the full connection amplitude becomes
\begin{equation}
 Q = Q_{0} \exp \left\{ -\frac{1}{k} \int^{\alpha}_{1} F(\nu)r^{\nu}d\nu \right\},
\label{eq15}
\end{equation}
where all possibilities from the line shape with $\nu = 1$
to another shape with $\nu = \alpha$ have been included.
The weight factor $F(\nu)$ has been introduced to account
for possible different contributions from different $\nu's$,
and the variable $r$ is
\begin{equation}
 r = \frac{1}{\ell} | \vec{x}- \vec{y} |
\label{eq16}
\end{equation}
with $\ell$ being a scale parameter. If we take $\alpha = 2$,
which corresponds to a spherical shape of in-hadron region,
and considering the case of equal weight $F(\nu) = 1$, we have~\cite{R10}
\begin{equation}
  Q = Q_{0} \exp \left \{ -\frac{1}{k} \frac{r^{2} - r }{\ln r} \right\}.
\label{eq17}
\end{equation}
This functional form describes smooth changes of the connectedness
of in-hadron region for large enough values of $r$, however,
for short ranges, we need to include perturbative local effects.
These effects can be accounted for if we substitute $r^{\beta}Q$ for $Q$,
in which case the conditions on $\mathfrak{M}$ still hold with $\beta > 0$.
Then the final form of $Q$ becomes
\begin{equation}
  Q = \frac{Q_{0}}{r^{\beta}} \exp \left\{ -\frac{1}{k} \frac{r^{2} - r }{\ln r} \right\}.
\label{eq18}
\end{equation}

%
%
\section{Estimation of nonlocal condensates}
%

With the above connection amplitude, we can estimate the values of
dimension 2 condensate $\langle A^{2}_{\mu} \rangle$ by assuming that
\begin{eqnarray}
 \langle :  \bar{q}(x)U(x,y)A^{a}_{\mu}(y)A^{\mu}_{a}(y)U(y,0)q(0) : \rangle \nonumber \\
 \propto \langle :  \bar{q}(x)U(x,y)q(y)\bar{q}(y)U(y,0)q(0) : \rangle ,
\label{eq19}
\end{eqnarray}
which implies that the probability amplitude to have a quark pair
is taken to be proportional to the value of $\langle A_{\mu}^{2} \rangle$ at that point.
For a meson with quarks at $\vec{r}_{1}$ and $\vec{r}_{2}$,
the condensate value at $\vec{r}$  becomes
\begin{equation}
\gamma(\vec{r}) = \gamma_{m}
\prod_{i=1}^{2} |\vec{r}-\vec{r}_{i}|^{-\beta}
\exp \left\{ -\frac{1}{k} \frac {|\vec{r}-\vec{r}_{i}|^{2}
- |\vec{r}-\vec{r}_{i}|}{\ln |\vec{r}-\vec{r}_{i}|} \right\}
\label{eq20}
\end{equation}
with an appropriate normalization factor $\gamma_{m}$.
If we fix the quark positions at $\vec{r}_{1} = (-\frac{r_0}{2}, 0)$
and $\vec{r}_{2} = (\frac{r_0}{2}, 0)$,
we can draw the values of dimension 2 condensate as in Fig.~\ref{fig1}.
We have changed the values of $\beta$ and $k$ and shown the results here
for the values $\beta=1.0$ and $k=1.0$.
For a baryon, the quark-to-quark connections are three fold
and therefore the condensate value at $\vec{r}$ becomes
%
\begin{eqnarray}
\gamma(\vec{r}) ~=~ \gamma_{b}  \prod_{i=1}^{3} |\vec{r}-\vec{r}_{i}|^{-\beta}
\exp \left\{ -\frac{1}{k} \frac {|\vec{r}-\vec{r}_{i}|^{2}
- |\vec{r}-\vec{r}_{i}|}{\ln |\vec{r}-\vec{r}_{i}|} \right\}~~  \nonumber  \\
\cdot \sum_{i=1}^{3}  \prod_{\vec{r}_{j}, \vec{r}_{k} \neq \vec{r}_{i}} |\vec{r}_{j}-\vec{r}_{k}|^{-\beta}
\exp \left\{ -\frac{1}{k} \frac { |\vec{r}_{j}-\vec{r}_{k}|^{2}
- |\vec{r}_{j}-\vec{r}_{k}|}{\ln |\vec{r}_{j}-\vec{r}_{k}|} \right\},
\label{eq21}
\end{eqnarray}
where $\vec{r}_{i}$ are the positions of three quarks.
The structure of dimension 2 condensates can be drawn as in Fig.~\ref{fig2}
with quark positions at $(0, 1)$, $(0, -1)$, and $(\sqrt{3}, 0)$.
We can see the Y-type connections which have been obtained in lattice calculations
through the estimate of gluon field components~\cite{R11}.
Of course the field components are squared to get the energy density or the action density,
however, the structures defined by constant densities
are not changed by squaring the components.

\begin{figure}[htb]
\begin{center}
\includegraphics[width=0.9\linewidth]{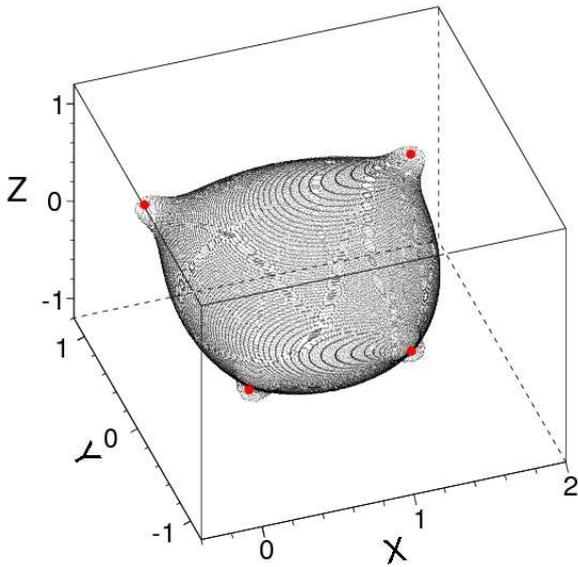}
\caption{ Equi-$\gamma$ surface for a tetraquark state with quarks at
$(0, 1, 0)$, $(0, -1, 0)$, $(\sqrt{2}, 0, 1)$, and $( \sqrt{2}, 0, -1)$.}
\label{fig3}
\end{center}
\end{figure}

\begin{figure}[htb]
\begin{center}
\includegraphics[width=0.9\linewidth]{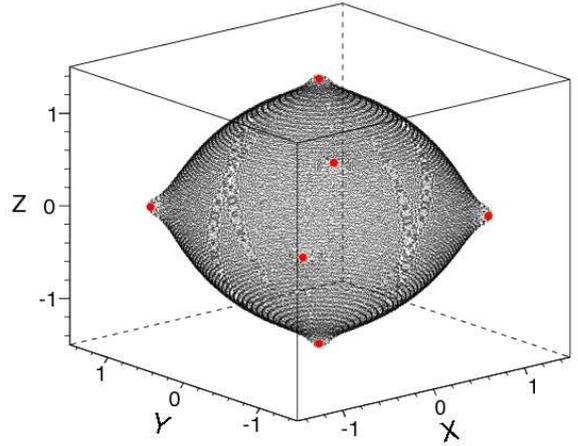}
\caption{ Equi-$\gamma$ surface for a hexaquark state with quarks at
$(0, 0, \sqrt{2})$, $(1, 1, 0)$, $(1, -1, 0)$, $(-1, -1, 0)$, $(-1, 1, 0)$,
and $(0, 0, -\sqrt{2})$.}
\label{fig4}
\end{center}
\end{figure}

The calculational techniques using the connection amplitude can be easily applied
to systems with complex boundary conditions such as multiquark states.
For tetraquark states, we have two possible combinations of color-singlet states
with one quark pair creation. One is the meson-tetraquark combination
and the other is the baryon-antibaryon combination.
The meson-tetraquark combination has $4$ amplitudes composed of $7$ connections
and the baryon-antibaryon combination has $1$ amplitude composed of $6$ connections.
The calculated results are shown in Fig.~\ref{fig3}
where only one surface with constant condensate value has been presented.
The quark positions are fixed at
$(0, 1, 0)$, $(0, -1, 0)$, $(\sqrt{2}, 0, 1)$, and $( \sqrt{2}, 0, -1)$.
As another example, we have calculated the case of hexaquark states
or deuteron for which we have $6$ amplitudes with $16$ connections
for the meson-hexaquark combination and $15$ amplitudes
with $13$ connections for the baryon-pentaquark combination.
The calculated results are shown in Fig.~\ref{fig4} with quark positions at
$(0, 0, \sqrt{2})$, $(1, 1, 0)$, $(1, -1, 0)$, $(-1, -1, 0)$, $(-1, 1, 0)$,
and $(0, 0, -\sqrt{2})$.
%

%
%
%

%
%
%
\section{Summary}
%

In summary, we have introduced the notion of in-hadron region
and constructed the topological spaces of in-hadron regions.
For a given space, it is possible to parameterize
the quark-to-quark connections with nonlocal measure
and to estimate the variations of dimension 2 condensate $\langle A_{\mu}^{2} \rangle$.
The estimated results are impressive for the systems of mesons and baryons
compared with the results of lattice gauge calculations.
Our method can be easily applied to systems with complex boundary conditions
and we have shown the surfaces of constant $\langle A_{\mu}^{2} \rangle$
for the systems of tetraquarks and hexaquarks.
The extension to the more complex nuclear states can be carried out straightforwardly
and we expect to get the picture of QCD vacuum for various nuclear states.
%
%
%
%
%

%
\end{document}